\documentclass[10pt,twocolumn]{article} 

\usepackage[letterpaper,margin=0.75in,columnsep=0.25in]{geometry}
\usepackage{times}
\usepackage[round]{natbib}   

\usepackage{amsmath,amsfonts,bm}

\def\eqref#1{equation~\ref{#1}}

\def\1{\bm{1}}

\DeclareMathAlphabet{\mathsfit}{\encodingdefault}{\sfdefault}{m}{sl}
\SetMathAlphabet{\mathsfit}{bold}{\encodingdefault}{\sfdefault}{bx}{n}

\usepackage{xcolor}    
\usepackage{hyperref}
\usepackage{url}
\usepackage{xurl} 
\hypersetup{colorlinks=true,linkcolor=blue!50!black,citecolor=blue!50!black,urlcolor=blue!50!black}
\usepackage{booktabs}
\usepackage{amsfonts}
\usepackage{amsmath}
\usepackage{amssymb}
\usepackage{graphicx}
\usepackage{microtype}

\title{metalSPARSE: A 2:4 Semi-Structured Sparse GEMV Kernel for Apple Silicon GPUs}

\author{Tek Raj Awasthi$^{1}$ \quad Babita Awasthi \\[4pt]
{\normalsize $^{1}$Friedrich-Alexander-Universit\"at Erlangen-N\"urnberg (cs.fau.de)}
}
\date{}   

\begin{document}

\maketitle

\begin{abstract}
Autoregressive Large language Models (LLMs) generate text one token at a time, and each token reads every weight from memory, so on edge devices decoding is limited by memory bandwidth. $2{:}4$ semi-structured sparsity keeps two of every four weights, but Apple-silicon GPUs have no sparse matrix hardware, so pruning helps only if a kernel actually reads fewer bytes. We present metalSPARSE, to our knowledge the first $2{:}4$ sparse kernels for Apple GPUs, together with a fused decoder written in Metal. The kernels stream weights in compressed form and expand them in registers. On Qwen2.5-1.5B with $2{:}4$ sparsity on every decoder linear layer, decoding runs $3.6\times$ faster than HuggingFace Transformers on PyTorch-MPS on an 8~GB M2 MacBook Air and $1.52\times$ faster than MLX on a 16~GB M4. The same engine on dense weights matches MLX, so the gain comes from the compressed weights, and sparsity delivers 97\% of the speedup its byte reduction allows. Prompt processing, which is compute-bound, does not speed up. To preserve quality, we combine a Wanda mask with block-wise reconstruction and low-rank adapters trained jointly with the surviving weights and quantized to 4 bits. On WikiText-2 this brings perplexity from $3.01\times$ dense after one-shot pruning (Wanda with a closed-form refit) to $1.34\times$, and the whole pipeline runs on the same laptop.
\end{abstract}

\section{Introduction}
\label{sec:intro}

Large language models increasingly run on the machine in front of the user. On Apple silicon that machine has a unified memory bus of 100 - 120~GB/s and a GPU with no sparse tensor cores, and single-stream decode (one token at a time, every weight read once) is bound by that bus. Every practical compression method for the platform therefore reduces bytes: 4- and 8-bit quantization, mixed precision, and structured pruning that exports a smaller dense checkpoint. Semi-structured $2{:}4$ weight sparsity, which halves the bytes of a matrix and is a first-class hardware feature on NVIDIA, AMD, and Meta accelerators, has been absent from this list for a structural reason. Without hardware support, a $2{:}4$ matrix stored densely still moves every zero, and the framework Apple ships, MLX, has declined to add a sparse-weight path; its sparse-matmul request is closed as \emph{wontfix} and a community SpMM kernel was rejected on surface-area grounds. The result is that a decade of $N{:}M$ pruning research has had no route to a Mac.

Two further obstacles compound the first. Post-training $2{:}4$ pruning is punishing at small scale: on Qwen2.5-1.5B the one-shot tier (Wanda selection with a closed-form per-matrix refit) leaves perplexity at $3.0\times$ dense. Block-wise reconstruction recovers much of this but not enough, and the strongest one-shot corrections, closed-form low-rank adapters as in SLiM, are derived for GPUs where the adapter's extra bytes and dispatches are nearly free. On a bandwidth-bound GPU with tens of microseconds per dispatch, an adapter that is not designed for the kernel it rides on can cost more than the bytes it saves; we measured exactly that before fixing it.

This paper closes the gaps from both sides. On the systems side, \textit{metalSPARSE} stores weights in a packed $2{:}4$ format (two fp16 values and four bits of metadata per block of four), streams them compressed, and expands each block in registers inside a family of Metal kernels tiered by batch size: a simdgroup-per-row GEMV for decode, register-tiled and \texttt{simdgroup\_matrix} variants for small batches, and P-GEMM, a prefill GEMM that decompresses $2{:}4$ tiles during the tile load and folds the low-rank correction into extra K-steps and the SwiGLU into its epilogue. The kernels sit inside a C++ decoder that owns all 28 layers with zero per-token allocations, merged command encoders, GQA-native attention with split-K decode, an INT8 tied head, and a fused decode MLP that runs a $2{:}4$ layer with rank-308 adapters in four dependent dispatches. The $2{:}4$ GEMV sustains 87.5~GB/s on an Mac air M2 whose measured streaming ceiling is 91~GB/s, and a matched-runtime decomposition (the same engine holding dense weights) shows that sparsity converts to time at 98\% of its byte ceiling. On dense weights the engine decodes within 1\% of MLX, which is what makes the sparse speedup attributable to the format rather than to runtime engineering.

On the modeling side, we assemble a recipe whose every stage is either measured to matter or measured to be replaceable. A Wanda mask selects the $2{:}4$ support; we show that once block-wise reconstruction follows, mask policy no longer matters (Wanda and SparseGPT masks converge to within 0.24\%, a statistical tie), so the cheap selection wins on Pareto grounds. Reconstruction quality is bound by calibration data rather than by optimizer sophistication, and scaling the calibration to the literature's token budget is the largest single one-shot gain. A low-rank correction, initialized in closed form and then trained jointly with the surviving values inside the block objective, beats the best closed form at matched bytes by 4.7\% on every one of 146 test windows; doubling its rank to 308 recovers 74\% of the one-shot tier's log-excess. On OPT-1.3B, under SLiM's own protocol, the same recipe comes within 2.3\% of SLiM-LoRA (1.2989 vs.\ 1.27) and decodes $4.45\times$ faster than HuggingFace eager (dense baseline model). The correction is folded into the dense attention weights for free and quantized to NF4 for the MLP, where it retains 98\% of its benefit: a correction term tolerates 13\% quantization error because that error is second-order in the loss. Every quality number is a paired, cluster-robust comparison on identical windows, and every speed number is decode-only milliseconds per token measured in situ, because we found that isolated kernel microbenchmarks invert format rankings and that syncing per-op profilers misprice ATen ops by up to $8\times$.

The shipped all-linear row ($2{:}4$ on every decoder linear layer, rank-308 INT4 adapters) reaches a paired WikiText-2 perplexity ratio of 1.34 and decodes at 25.3~ms/token on a fanless 8~GB M2 ($3.6\times$ PyTorch-MPS eager) and 21.2~ms/token on an M4 ($1.52\times$ MLX, $1.55\times$ the oMLX server). The MLP-only hybrid row, which prunes 24 of 28 MLPs and leaves attention dense, reaches 1.22 at $3.6\times$ decode and $1.16\times$ prefill on the M2. We state the ceiling plainly: prefill is compute-bound, $2{:}4$ removes no FLOPs on this hardware, and the all-linear row's prefill sits at $0.73$–$0.94\times$ dense, the disclosed cost of its adapters.

Our contributions are:
\begin{itemize}
    \item \textbf{The first $2{:}4$ semi-structured sparse kernel family and fused decoder for Apple GPUs} (to our knowledge): batch-tiered GEMV kernels, a decompress-in-tile-load prefill GEMM with adapter K-extension, 4-bit metadata, and a zero-allocation C++ decoder. The GEMV runs at 96\% of measured STREAM bandwidth; sparsity realizes 98\% of its byte ceiling; the runtime matches MLX on dense weights.
    
    \item \textbf{A laptop-scale $2{:}4$ recipe and its ablations}: Wanda mask -\textgreater{} block-wise reconstruction with scaled calibration -\textgreater{} jointly trained low-rank correction -\textgreater{} NF4 adapters. We show mask policy is irrelevant after block reconstruction, calibration volume is the binding one-shot constraint, joint training beats closed-form corrections at matched bytes (52~SE), and 4-bit adapters keep 98\% of their benefit at two ranks.
    
    \item \textbf{Kernel-level results for quantized corrections on Apple GPUs}: a quantized adapter's cost is its memory-access pattern, not its format: our first kernels lost 9.5~ms/token streaming a quarter of the bytes; the rewrite realizes the byte model exactly on two GPU generations.
    
    \item \textbf{A measurement methodology for unified-memory inference}: matched-runtime decomposition (runtime $\times$ head $\times$ format), paired cluster-robust quality instruments with measured design effects, in-situ kernel pricing, and fourteen platform findings including a silent MPS matmul defect above 1024 rows, allocator heap granularity that doubles resident memory, and thermal and determinism rules.
    
    \item \textbf{An open release}: kernels, engine, pruned models, benchmark harness, and the negative results.
\end{itemize}

\section{Related Work}
\label{sec:related_work}

\paragraph{Post-training pruning of LLMs.}
One-shot pruning decomposes into mask selection and value reconstruction. SparseGPT \citep{sparsegpt} selects greedily inside a compensated OBS sweep; Wanda \citep{wanda} selects a priori from $|W|\cdot\|X\|$ and updates nothing; OWL \citep{owl} and DSnoT \citep{dsnot} refine the layer budget and the mask after the fact. Reconstruction has since been separated from selection and strengthened: ALPS \citep{alps} solves the layer-wise problem with ADMM, Wanda++ \citep{wandapp} optimizes each decoder block regionally, and \citet{freelunch} show that block-wise reconstruction dominates per-matrix reconstruction and that once it is applied, differences between selection criteria shrink with model size. The scaling of recovery with post-training data is studied by \citet{p2law}. We adopt Wanda selection and block-wise reconstruction and extend these findings downward, to 1.5B parameters and the $2{:}4$ constraint on every linear layer: mask policy becomes statistically irrelevant after block reconstruction, and calibration volume, not optimizer sophistication, is the binding constraint on one-shot quality.

\paragraph{Semi-structured sparsity and its hardware.}
$N{:}M$ sparsity was introduced with NVIDIA's sparse tensor cores \citep{ampere24} and is now supported on AMD GPUs and Meta's MTIA. Learned-mask methods close much of the $2{:}4$ quality gap at training-scale cost: MaskLLM \citep{maskllm} learns masks end to end, ProxSparse \citep{proxsparse} and \citet{proxop} regularize toward the pattern, SparseForge \citep{sparseforge} anneals a Hessian-guided soft mask, and AST \citep{ast} retrains with distillation. Relaxed formats keep hardware compatibility while easing the 50\% constraint: PATCH \citep{patch} mixes dense and sparse tiles, SlideSparse \citep{slidesparse} expands activations to express $(2n{-}2){:}2n$, and SpenseGPT \citep{spensegpt} keeps a saliency-chosen dense region of each MLP and prunes attention not at all. All of this work assumes a sparse GEMM that the hardware provides. We target a GPU that provides none, and the only lever left there is bytes; our MLP-only hybrid row is a layer-level instance of Spense's dense region, and index-level selection on our kernels is future work.

\paragraph{Sparse, low-rank, and quantized compression.}
Low-rank terms have been used to absorb compression error in several forms: LoSparse \citep{losparse} decomposes $W \approx UV + S$, LQ-LoRA \citep{lqlora} and RoSA \citep{rosa} train adapters over quantized or sparse bases at fine-tuning cost, and L$^2$QER, QERA and CALDERA \citep{l2qer,qera,caldera} derive closed-form adapters for quantization error. SLiM \citep{slim} is closest to us: it combines per-tensor quantization, Wanda $2{:}4$ selection, and a closed-form saliency-weighted low-rank correction, quantizes the adapters to 4 bits with group AbsMax, and reports that they lose little accuracy. JSQ \citep{jsq} and \citet{harma2024} study the ordering and interplay of sparsity and quantization. We differ in three ways. Our correction is initialized in closed form and then trained jointly with the surviving $2{:}4$ values inside the block objective, which beats the best closed form at matched bytes on every test window. We quantify \emph{how much} of a correction's benefit survives 4-bit quantization across two ranks (98\%) and explain it as a second-order effect, and we show that the details that matter are an activation-weighted clip and an NF4 codebook for the Gaussian left factor. And we design the kernels the adapters run on, folding the attention adapters into dense weights at no cost and streaming the MLP adapters through a fused epilogue.

\paragraph{Inference on Apple silicon.}
Compression on Apple silicon is quantization: MLX \citep{mlx} and llama.cpp \citep{llamacpp} ship affine 4- and 8-bit formats, K-quants, and MXFP4/MXFP8, and \citet{llmflash} treats the flash-to-DRAM boundary. Native Metal runtimes such as vllm-mlx \citep{vllmmlx} and BaseRT \citep{basert} report 1.2--1.6$\times$ over framework runtimes from kernel fusion and dispatch elimination alone. No runtime has a $2{:}4$ weight-sparse path: MLX's sparse-matmul request is closed as \emph{wontfix} and a contributed CSR kernel was declined on surface-area grounds\footnote{Issue and PR numbers to be verified against the repository at camera-ready.}; \texttt{mlx-sparse} provides COO/CSR arrays but is not wired into any linear layer; Apple's MPSGraph exposes general sparse-tensor ops and MPS exports an undocumented sparse matmul symbol, neither $N{:}M$. Sparse GEMM on the platform exists on the CPU via NEON \citep{ternaryapple}, sparse linear algebra in Metal exists for cryptographic workloads, and Metal sparse-\emph{attention} kernels \citep{opentq} sparsify no weight matrix. To our knowledge metalSPARSE is the first $2{:}4$ weight-sparse kernel family for the Apple GPU. Because runtime engineering alone is worth a known $1.5\times$, we hold it fixed: our engine on dense weights decodes at parity with MLX, so the speedup we report is the format's.

\subsection{The Memory Bandwidth Wall on Edge Platforms}
\label{sec:bandwidth-wall}

\paragraph{Decode is a bandwidth problem.}
A kernel performing $F$ FLOPs over $B$ bytes, on a device with peak
arithmetic $\pi$ and sustained bandwidth $\beta$, is memory-bound when its
arithmetic intensity $F/B$ lies below the ridge point $\pi/\beta$, and then
takes at least $B/\beta$ \citep{williams2009roofline,yuan2024llmviewer}.
Decoding one token reads every weight once for a single multiply--add, an
intensity of ${\approx}1$~FLOP/byte in fp16. Servers amortize those reads by
batching requests \citep{pope2023scaling}; an edge device serves one user and
cannot. Qwen2.5-1.5B \citep{qwen25} streams 3.09~GB of fp16 weights per token
for 3.09~GFLOP. On our 8-core M2 GPU (2.86~TFLOP/s fp16 peak, 91~GB/s measured
streaming bandwidth, ridge ${\approx}31$~FLOP/byte), the arithmetic needs
1.1~ms and the reads at least 33.9~ms: decode time is bytes over bandwidth.

\paragraph{The wall is widening.}
Server compute has grown $3.0\times$ every two years against $1.6\times$ for
DRAM bandwidth \citep{gholami2024memorywall}, and edge SoCs follow the same
curve. Bandwidth of Apple's base M-series rose from 100~GB/s
(M2; \citealp{applem2}) to 120 (M4) and 153~GB/s (M5), while the M5 added a
dense matrix unit to every GPU core: Apple reports up to $4\times$ faster
time-to-first-token than the M4 but only 19--27\% faster generation, tracking
the 28\% bandwidth gain \citep{applemlxm5,basertm5}. Each generation makes
decode more bandwidth-bound, not less. Capacity is the second wall: CPU, GPU
and OS share one LPDDR pool, and on our 8~GB machine Metal recommends a GPU
working set of at most 5{,}461~MiB (two-thirds of physical memory), 54\% of
it taken by the fp16 model alone; exceeding it collapses throughput
(\S\ref{sec:memory}).

\paragraph{What 2:4 can buy on an Apple GPU.}
Metal's matrix primitive, \texttt{simdgroup\_matrix}, multiplies dense
$8{\times}8$ tiles, the M5's accelerators also take dense operands, and no
Apple GPU exposes structured-sparse matrix instructions \citep{metalspec}.
2:4 therefore pays through bytes alone: fp16 values plus 2-bit indices take
1.125~B per weight slot, $\rho=0.5625$ of dense fp16. Three consequences
follow.
(i)~\emph{Amdahl's law over bytes.} If component $i$ streams $b_i$ bytes per
token at storage ratio $\rho_i$, the format alone speeds up decode by at most
\begin{equation}
  S_{\mathrm{fmt}} \;\le\; \frac{\sum_i b_i}{\sum_i \rho_i\, b_i}.
  \label{eq:byte-ceiling}
\end{equation}
The tied 151{,}936-token LM head is 15\% of dense traffic, so 2:4 on every
decoder linear with an fp16 head is capped at $1.59\times$; an INT8 head
raises the cap to $1.81\times$. Quantization \citep{mlx,llamacpp} lowers
$\rho_i$ through bit-width and 2:4 through count, so the two compose. Any
speedup beyond Eq.~\eqref{eq:byte-ceiling} comes from the runtime, which is
why we decompose every speedup we report (\S\ref{sec:decomposition}).
(ii)~\emph{Weights dominate traffic at edge context lengths:} with
grouped-query attention the KV cache adds 28~KiB per context token (1.9\% of
fp16 weight traffic at 2{,}048 tokens) and would overtake the 2:4 weights
only beyond ${\sim}60$K tokens, past the model's 32K native context.
(iii)~\emph{Dense-tile kernels cannot accelerate prefill.} Over $M$ prompt tokens a
linear layer's intensity is ${\approx}M$~FLOP/byte for dense fp16 and
${\approx}1.78M$ for 2:4 multiplied densely after decompression, so on the M2
the two turn compute-bound beyond ${\approx}31$ and ${\approx}18$ tokens
respectively. Past that point the byte saving is invisible and decompression
is overhead. Halving the multiply work instead requires a kernel that skips the
zeros,\footnote{On M1--M4, \texttt{simdgroup\_matrix} runs on the same FP32 ALUs
  as scalar code \citep{rigel}, so zero-skipping would face no hardware penalty
  for leaving the matrix path; the cost is operand delivery.} and then every
surviving weight must fetch its activation by index, forfeiting the operand
reuse of \texttt{simdgroup\_matrix} tiles. Measured at 512 tokens on the MLP
shapes (Table~\ref{tab:zero-skip}), a register-blocked zero-skipping kernel
reaches 0.82--0.86 dense-equivalent TFLOP/s, against 1.81--1.82 for our
dense-tile 2:4 GEMM (\S\ref{sec:kernel_design}) and 2.15--2.20 for the MPS
dense GEMM; skipping zeros makes it only $1.14\times$ faster than its own dense
twin, not $2\times$. The prefill ceiling is therefore parity, and any prefill
gain must come from the runtime, unlike on GPUs whose sparse tensor cores
halve the multiply work in hardware \citep{ampere24}.

\section{The metalSPARSE Method: Recipe, Kernels, and Engine}
\label{sec:kernel_design}

\begin{figure*}[t]
\centering
\includegraphics[width=\linewidth]{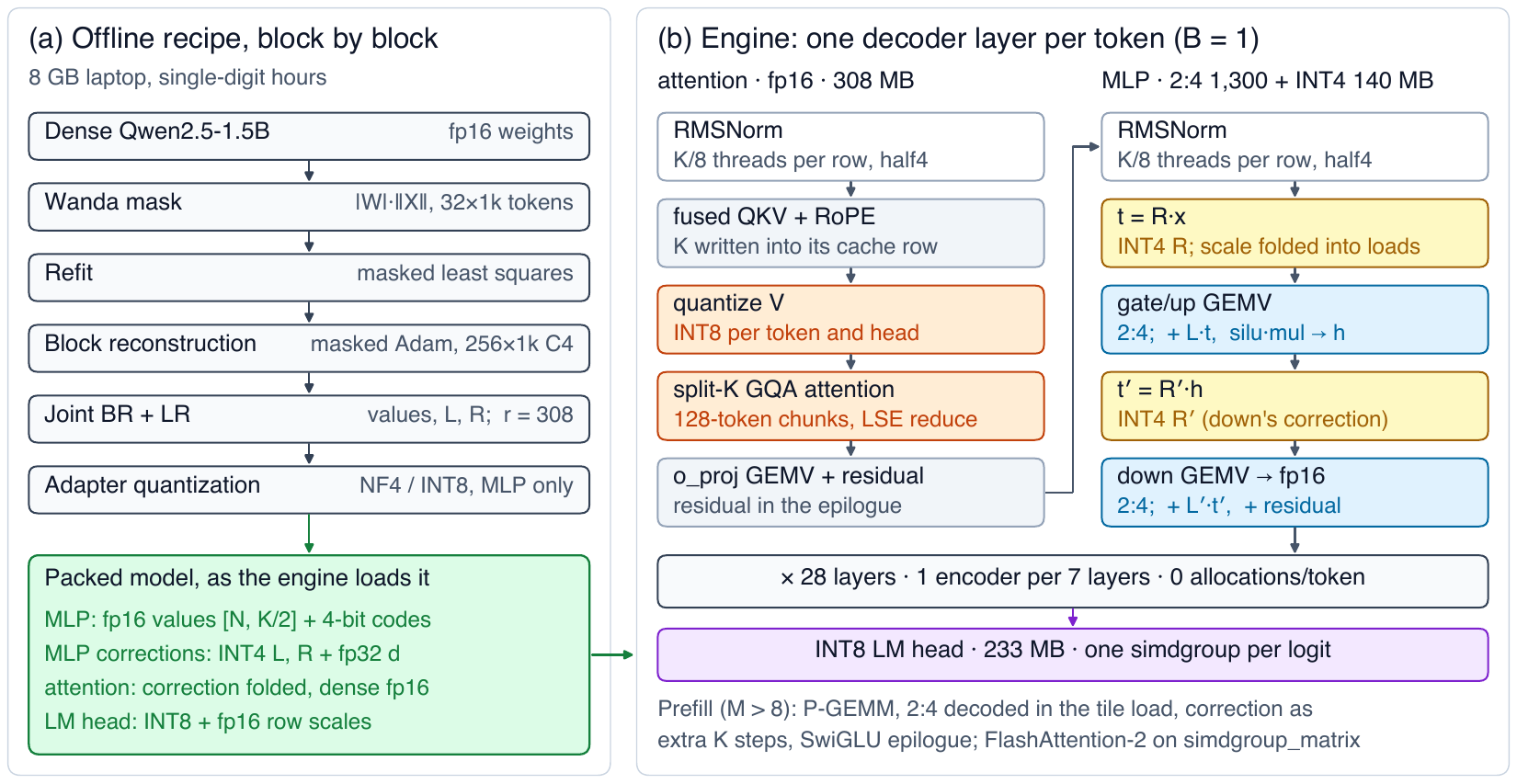}
\caption{The two halves of metalSPARSE meet at a packed file. (a) The offline
recipe (\S\ref{sec:recipe}) and the storage format of every tensor in the file
it writes. (b) One decoder layer of the engine at decode
(\S\ref{sec:kernels}, \S\ref{sec:engine}); colour marks the format each kernel
streams: blue 2:4, yellow INT4 correction, orange INT8 V cache, purple INT8
head, grey dense fp16. Attention is dense fp16 with its correction folded in;
the MLP is four dependent dispatches. Sizes are bytes per token over all 28
layers.}
\label{fig:pipeline}
\end{figure*}

The system has two halves that meet at a packed file (Fig.~\ref{fig:pipeline}):
an offline recipe decides which weights survive, what their values are and how
the correction that recovers quality is stored; an online C++/Metal engine
streams those bytes at the machine's bandwidth. We ship two configurations.
\textbf{50\%-all-linear}: 2:4 on all 196 decoder linear layers, C4-calibrated,
a rank-308 correction on every layer, MLP corrections in 4~bits and attention
corrections folded into the dense weights; 42.4\% of all parameters removed.
\textbf{32\% hybrid}: 2:4 on the MLPs of 24 of 28 layers, attention dense, no
correction; blocks 0, 1, 26 and 27, the most sensitive to weight removal, stay
dense; 32.1\% of parameters, i.e.\ 37.8\% of decoder-linear weights,
removed, the figure to use against methods that remove 50\%.

\subsection{The pruning recipe}
\label{sec:recipe}
\textbf{Objective.} For decoder block $b$ with a fixed 2:4 mask $M$, every stage
minimises the block's own output error over calibration inputs $x$,
\begin{equation}
  \min_{W\odot M,\;L,R}\;
  \big\|\mathrm{MLP}_b\big(x;\,(W\odot M)+LR\big)-\mathrm{MLP}_b^{\mathrm{dense}}(x)\big\|_2^2,
  \label{eq:block-objective}
\end{equation}
where $W$ ranges over the block's gate, up and down matrices, $M$ keeps two of
every four weights along the input dimension, and the rank-$r$ pair $(L,R)$ is
the correction, absent until the joint stage. Blocks are processed
sequentially: block $b$ is fed inputs from the model with blocks $1\ldots b{-}1$
already compressed, and its target comes from its own still-dense weights
before they are overwritten, so upstream error is propagated, never hidden.
In the all-linear row the attention projections take the mask, the refit
values and, later, their closed-form correction without a gradient step: a
lone matrix has no block objective.

\textbf{Mask.} Wanda's score $S_{ij}=|W_{ij}|\,\|X_j\|_2$ keeps the two largest
of every four weights along the input dimension, from activation norms over 32
sequences of 1{,}024 tokens, computed once and frozen. We do not use
SparseGPT's OBS selection: once block reconstruction follows, the two masks
tie (0.24\%, 0.7~SE; \S\ref{sec:quality_recovery}), so the cheaper mask wins.

\textbf{Refit (warm start).} Per output row, the masked least squares that
reproduces the dense weights on the already-pruned input,
$w_S=(H_{SS}+\lambda I)^{-1}H_{S,:}\,w^{\mathrm{dense}}$, with
$H=X^{\!\top}X$ on the propagated input, $S$ the row's kept support and
$\lambda$ one percent of the mean diagonal. It initialises the next stage.

\textbf{Block reconstruction.} Masked Adam on Eq.~\eqref{eq:block-objective}
without $LR$: fp32 working copies, learning rate $2{\times}10^{-4}$, and a
re-projection onto $M$ after every step so pruned positions stay exactly
zero. The calibration rule scales \emph{unique} data at constant epochs: 256
sequences of 1{,}024 C4 tokens (240 train, 16 held out), 2{,}400 steps (ten
epochs) with per-epoch shuffling, a 10-step warm-up and cosine decay to 10\%;
held-out error is sampled every 25 steps and the best point kept, and a
do-no-harm gate keeps the initialisation whenever held-out error did not
improve. The recipe runs on an 8~GB laptop in single-digit hours.

\textbf{Joint reconstruction with a low-rank correction.} Each matrix becomes
$y=(W\odot M)x+LRx$. The pair starts from SLiM's closed form: with residual
$E=W^{\mathrm{dense}}-W$ and $s_j^2=H_{jj}+\lambda$,
$LR=\mathrm{SVD}_r(E\,\mathrm{diag}\,s)\,\mathrm{diag}(s)^{-1}$; values and both
factors then train jointly on Eq.~\eqref{eq:block-objective}, the gate
comparing against this adapted initialisation. The shipped rank is 308, 20\%
of the hidden size; ``low-rank'' is being stretched, and the next stage is
what makes it affordable. Attention corrections are folded into the attention
weights at load time, a measured choice: unfolding them to pack attention 2:4
would save about 73~MB/token but cost 0.6--0.9~ms of dispatches, a net of
roughly zero. Attention is therefore 2:4-constrained for quality but streamed
dense.

\textbf{Adapter quantization.} MLP corrections only, after training; the 2:4
values stay fp16. With $LR=L_u\,\mathrm{diag}(d)\,R_u$, $L_u$ is NF4 with one
scale per output row (its entries are near-Gaussian with a flat spectrum),
$R_u$ is stored in groups of 128 along the input with an activation-aware
column scale $c_j=(s_j/\mathrm{median}\,s)^{\alpha}$, $\alpha$ searched per
matrix, and $d$ stays fp32 and is applied inside the kernel. The clip search
is weighted by $s^2$; an unweighted clip removes exactly the entries the
activations weight most (error 14.9\% $\to$ 18.7\%). A consistency contract
keeps the numbers honest: every fp16 consumer (evaluation, prefill, the dense
mirror) uses the dequantized factors, verified per layer against the packed
buffers, so the quality instruments describe the model the decode path runs.
The pre-registered rule ships the smallest format within 0.5\% of the fp16
correction on the paired WikiText-2 gate with a clean sink probe; at rank 308
the 4-bit format exceeds the bar by 0.15 points and ships as a disclosed
override (\S\ref{sec:quality_recovery}).

\subsection{The 2:4 kernels}
\label{sec:kernels}
\begin{figure*}[t]
\centering
\includegraphics[width=\linewidth]{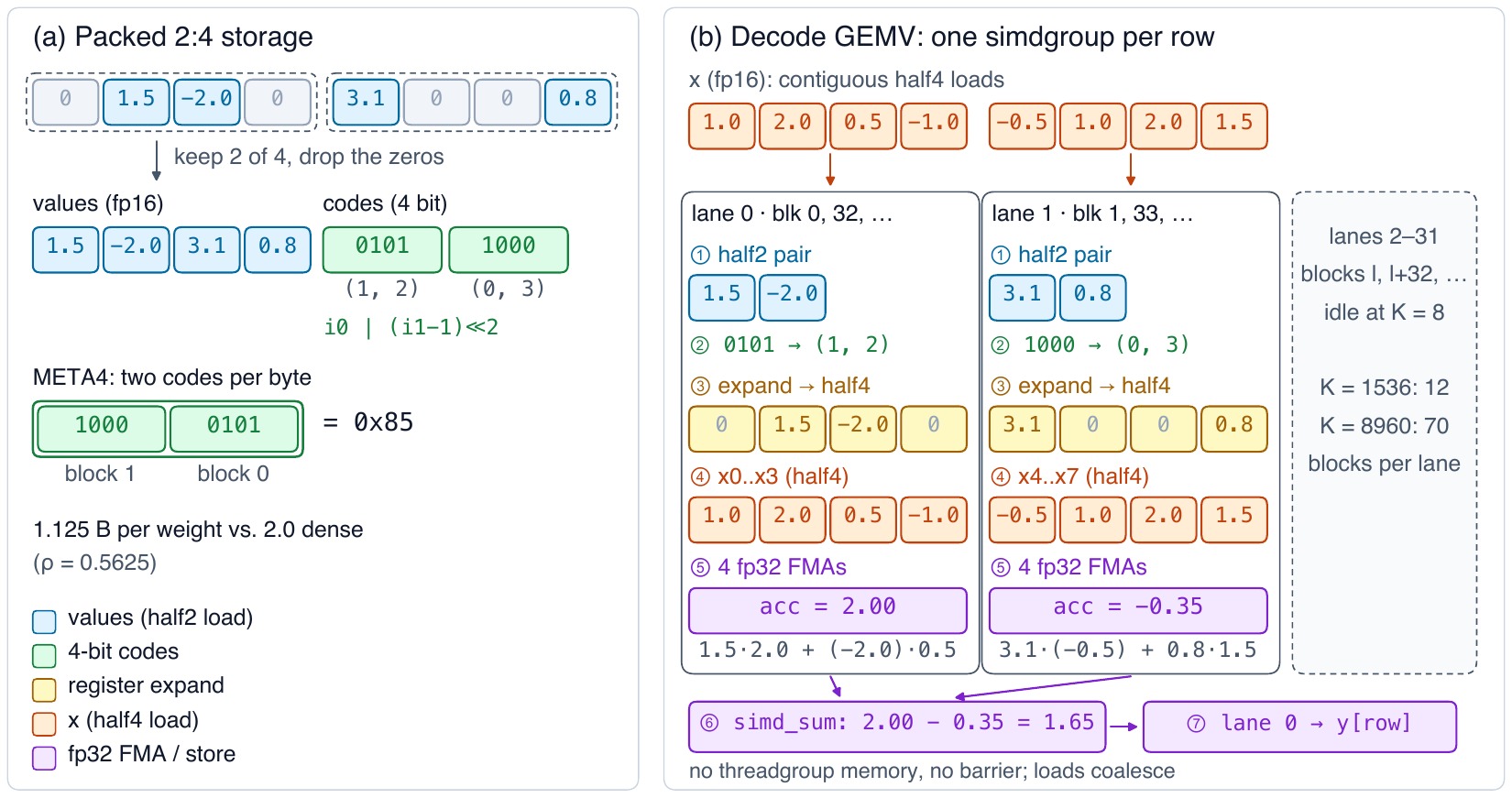}
\caption{(a) The packed 2:4 format: the two kept weights of each block in
position order and one 4-bit code, $\texttt{i0}\,|\,(\texttt{i1}{-}1)\ll 2$,
per block (two per byte); 1.125~B per weight slot. (b) The decode GEMV: lane
$l$ of the row's simdgroup owns blocks $l, l{+}32,\ldots$; each block is one
\texttt{half2} load, a register expansion to a \texttt{half4} with its zeros,
one contiguous \texttt{half4} load of $x$ and four fp32 FMAs; \texttt{simd\_sum}
reduces the lanes and lane 0 stores. Numbers follow one row with $K{=}8$.}
\label{fig:format-gemv}
\end{figure*}

\textbf{Format.} $W\in\mathbb{R}^{N\times K}$ is stored as the two kept weights
of each block of four, in position order, as fp16 values $[N,K/2]$, plus one
4-bit code per block, $\texttt{i0}\,|\,(\texttt{i1}{-}1)\ll 2$ for kept
positions $\texttt{i0}<\texttt{i1}$, two codes per byte: 1.125~B per weight
slot, $\rho=0.5625$ of dense fp16 (Fig.~\ref{fig:format-gemv}a). Decoding a
block is one 32-bit \texttt{half2} load, one nibble, and a scatter of the pair
into a zero-initialised \texttt{half4} in registers, never a dense copy.

\textbf{Decode GEMV, tiered by batch.} At $B\leq 3$ the grid is one 32-lane
simdgroup per (row, batch item); lane $l$ owns blocks $l, l{+}32,\ldots$ and per
block loads its pair, expands it, loads the \emph{contiguous} \texttt{half4} of
$x$ at that column and issues four fp32 FMAs (Fig.~\ref{fig:format-gemv}b).
Adjacent lanes read adjacent words, so both loads coalesce; there is no
threadgroup memory and no barrier; \texttt{simd\_sum} reduces the lanes and
lane~0 stores. Two of the four FMAs multiply a zero, deliberately: at decode
the kernel is bound by the bytes it streams and the expansion keeps every
load contiguous (\S\ref{sec:bandwidth-wall} shows why skipping zeros loses
when compute-bound). The MLP chain runs at 87.5~GB/s, 96\% of the M2's STREAM
bandwidth, and 81 function-constant shapes of the kernel land within $-1.2$ to
$+1$\% of it. At $B=4$--$7$, $x$ is transposed once so one \texttt{half4} load
serves four batch rows; at $B=8$ each lane expands one block into
$8{\times}32$ tiles for four $8{\times}8$ \texttt{simdgroup\_matrix}
multiply-accumulates per $K$ step. Verifying four draft tokens costs 0.99 of
one token, the basis of our multi-row and batched decoders; above $B=8$ the
GEMV tiers lose to a GEMM.

\textbf{P-GEMM: prefill on \texttt{simdgroup\_matrix}.} At 514 prompt tokens
the M2 is compute-bound, and P-GEMM is one kernel with three
function-constant modes over a $64{\times}64$ output tile with four
simdgroups and fp32 accumulation; its dense core reaches 94--96\% of the MPS
GEMM at the prefill shapes after three tile-loop changes: rasterising
threadgroups $M$-fastest so a weight tile is read about once, skipping the
MMAs of tail tiles, and prefetching the next $K$ step's tiles into registers.
\textsc{Sparse} decodes 2:4 blocks while the weight tile is staged, so there
is no decompression pass and no dense scratch; \textsc{Dual} keeps gate and
up accumulators for one $X$ tile and applies $h=\mathrm{silu}(g)\cdot u$ in the
epilogue; \textsc{Ext} runs the correction $TL^{\!\top}$, $T=XR^{\!\top}$, as
$r$ extra $K$ steps in the same accumulators. An MLP layer is four dispatches
(eleven before); at rank 308 it matches the previous MPS-based path in time
and is $2.5$--$2.7\times$ closer to fp32, since the adapter sums and the SwiGLU
stay in fp32.

\textbf{The decode MLP in four dispatches.} A single kernel for the whole MLP
is not buildable on Metal: the down projection needs all of $h$, and Metal
has neither a grid-wide barrier nor a forward-progress guarantee for a spin
barrier, so fusion stops at that dependency: one kernel computes gate row $n$
and up row $n$ from one read of $x$, adds the correction for row $n$, applies
silu$\cdot$mul and writes $h$; the projection $t=Rx$ folds its per-channel
scale into its loads; the down GEMV's epilogue adds its correction, the
residual and the fp16 conversion. A rank-308 layer runs in \textbf{four}
dependent dispatches (ten before; two without a correction), exact to
summation order, so the quality instruments, which run the prefill path, are
untouched.

\subsection{The decode engine, by the terms of Eq.~\eqref{eq:byte-ceiling}}
\label{sec:engine}
Every remaining lever lowers a component's $\rho_i$ (bytes) or brings a
kernel's efficiency $\eta$ toward STREAM. Table~\ref{tab:bytes} is the byte
budget that makes Eq.~\eqref{eq:byte-ceiling} concrete; the kernels that
stream it are priced in Table~\ref{tab:kernels}.

\begin{table*}[t]
\caption{Bytes streamed per decoded token (MB, 512-token context), from the
engine's buffer sizes. The 2:4 rows carry 4-bit metadata, an INT8 head and an
INT8 V cache; the 50\% row adds rank-308 INT4 corrections.}
\label{tab:bytes}
\centering
\begin{tabular}{lrrr}
\toprule
Component & dense fp16 & 32\% hybrid & 50\% all-linear \\
\midrule
MLP weights (28 layers)        & 2{,}312 & 1{,}445 & 1{,}300 \\
Attention $q,k,v,o$            & 308     & 308     & 308 \\
LM head (tied)                 & 467     & 233     & 233 \\
Low-rank corrections           & --     & --     & 140 \\
KV cache at 512 tokens         & 14      & 11      & 11 \\
\midrule
Total                          & 3{,}101 & 1{,}997 & 1{,}992 \\
$\rho$ vs.\ dense              & 1.00    & 0.644   & 0.642 \\
Floor at 91~GB/s (ms)          & 34.1    & 21.9    & 21.9 \\
Measured decode, M2 (ms)       & 90.5$^{\dagger}$ & 24.8  & 25.3 \\
\bottomrule
\end{tabular}\\[2pt]
$^{\dagger}$PyTorch eager, dispatch-bound; the matched-runtime dense engine is
decomposed in \S\ref{sec:evaluation}.
\end{table*}

\textbf{Format levers ($\rho$).} 4-bit metadata, $-145$~MB/token and
bit-identical on every path; the INT8 tied head with one fp16 scale per row,
467 $\to$ 233~MB; an INT8 V cache per token and head, dequantized in
registers; 4-bit corrections, 543 $\to$ 140~MB at rank 308. The 50\% row
streams 1{,}992~MB in 25.3~ms, 78.7~GB/s or 86\% of STREAM; the hybrid row
streams 1{,}997~MB in 24.8~ms, 89\%.

\textbf{Kernels at the bandwidth ceiling ($\eta$).} Beyond the 2:4 GEMV: a
fused QKV+RoPE GEMV that writes $K$ into its cache row (85--89~GB/s); an
$o$-projection GEMV with the residual in its epilogue (82.5); the INT8 head,
one simdgroup per logit with \texttt{char4} loads and no threadgroup memory
(71.6); GQA-native attention that reads the shared KV head and runs an fp32
online softmax split over 128-token context chunks with a log-sum-exp reduce
(context cost 4.9 $\to$ 1.2~ms per 1{,}000 tokens); and the INT4/NF4 correction
kernels, whose rule is that \emph{cost is access pattern, not format}: one
4-byte load carries eight elements, $x$ is pre-scaled once per token, $L$ is
stored block-transposed so each expansion step is one coalesced load, and the
rank is padded to a multiple of eight so no kernel has a tail branch. Our
first version broke that rule and lost 9.45~ms/token while streaming a quarter
of the bytes (\S\ref{sec:evaluation}). Prefill uses a FlashAttention-2 kernel
on \texttt{simdgroup\_matrix} (0.96 vs.\ 3.78~ms per layer at 514 tokens for
\texttt{repeat\_kv} + MPS SDPA) and P-GEMM.

\textbf{Exact folding and fewer dispatches.} Attention corrections fold into
the weights, the per-token V scale into the softmax weights, and residual,
bias, RoPE, fp16 conversion and SwiGLU into epilogues. With no ATen op inside
a layer, decode runs in one compute encoder per commit window instead of
about thirteen per layer, committed without blocking every seven layers:
bit-identical and 1.9--2.5~ms/token faster. The decoder is a C++ loop over all
28 layers with a preallocated KV pool, a static pipeline-state cache and zero
allocations per token; the framework, not the kernel, was the first
bottleneck (a $2.9\times$ kernel ran at $0.01\times$ inside HuggingFace decode).

\textbf{Memory residency.}
\label{sec:memory}
Metal recommends a GPU working set of two-thirds of physical memory
(5{,}461~MiB on 8~GB); exceeding it pages the weights decode is streaming, and
with a dense prefill mirror resident the M2 fell from $2.9\times$ to
$0.84\times$. The engine budgets against Metal's own counters, allocates
persistent tensors as slabs outside the allocator's 1~GiB heaps, and keeps
per-layer metadata in few buffers, since residency cost follows buffer count,
not bytes (\S\ref{sec:evaluation}). The mirror, which lifts prefill to parity,
is a memory-for-latency mode for the 16~GB machine only.

\section{Quality Recovery: Beyond One-Shot Pruning}
\label{sec:quality_recovery}

\textbf{Instrument.} Every quality number is a perplexity ratio to dense on the
WikiText-2 test split in 146 non-overlapping 2{,}048-token windows (298{,}862
scored tokens; dense 9.265). Each arm is first checked against its unfused
fp32 CPU forward (within $4{\times}10^{-4}$ nats), and the LM head runs in
chunks of at most 1{,}024 rows, which keeps the MPS defect of
\S\ref{sec:hardware_defect} unreachable. Error bars are window-clustered. Two
methods are compared per token on identical windows, where shared difficulty
cancels: paired standard errors are 0.08--0.09\%, against 0.5--0.9\% for a
reconstructed row's ratio to dense. The 50\% row is calibrated on C4 and evaluated on WikiText-2,
SLiM's protocol \citep{slim}; the hybrid row is calibrated in distribution on
WikiText-2, which is worth 0.12~nats at its one-shot tier. Table~\ref{tab:quality}
is the ladder of both rows.

\begin{table*}[t]
\caption{Quality ladder, WikiText-2 test/2048 (ratio to dense, $\pm$ clustered
SEM). \emph{Recovered}: share of the one-shot tier's log-excess removed,
$1-\ln r/\ln r_{\text{one-shot}}$, the axis on which rows of different models
compare. \emph{MB}: correction bytes streamed per token by the engine (MLP;
attention corrections are folded).}
\label{tab:quality}
\centering
\small
\begin{tabular}{lrrr}
\toprule
Stage & Ratio & Recovered & MB \\
\midrule
\multicolumn{4}{l}{\emph{50\% all-linear, C4-calibrated}} \\
Wanda mask + refit (one-shot tier)       & 3.0103 $\pm$2.40\% & --    & -- \\
+ block reconstruction, c256             & 1.9124 $\pm$0.87\% & 41.2\% & -- \\
+ closed-form correction, $r{=}154$      & 1.5909 $\pm$0.68\% & 57.9\% & 272 \\
+ joint training, $r{=}154$              & 1.5166 $\pm$0.66\% & 62.2\% & 272 \\
+ joint training, $r{=}308$              & 1.3335 $\pm$0.53\% & 73.9\% & 543 \\
\quad correction in INT8                 & 1.3336 $\pm$0.53\% & 73.9\% & 275 \\
\quad \textbf{correction in INT4 (shipped)} & \textbf{1.3422} $\pm$0.54\% & \textbf{73.3\%} & \textbf{140} \\
\midrule
\multicolumn{4}{l}{\emph{32\% hybrid, WikiText-2-calibrated, no correction}} \\
Wanda mask + refit (one-shot tier)       & 1.5152 $\pm$0.72\% & --    & -- \\
+ block reconstruction, c32              & 1.3554 $\pm$0.65\% & 26.8\% & -- \\
+ block reconstruction, c128             & 1.2502 $\pm$0.55\% & 46.3\% & -- \\
+ block reconstruction, c256 (shipped)   & \textbf{1.2167} $\pm$0.54\% & \textbf{52.8\%} & -- \\
\bottomrule
\end{tabular}
\end{table*}

\textbf{Block reconstruction: granularity first, then data.} Replacing the
per-matrix refit with the block objective lowers the 50\% row's ratio by
36.5\% (21.4~SE) and wins every window. The gain is the granularity, not the
mask: SparseGPT and Wanda masks enter block reconstruction from
initialisations 4.0\% apart and leave it 0.24\%~$\pm$~0.36\% apart (0.7~SE, validation windows), so
the mask that costs seconds replaces an OBS sweep of about 40 minutes. The
next-largest lever is calibration volume. Scaling \emph{unique} data at a
constant ten epochs, 32 $\to$ 128 $\to$ 256 sequences of 1{,}024 tokens,
lowers the hybrid row by 7.8\% and then 2.7\%, each paired; on validation
windows the same recipe gives the same step on the SparseGPT mask ($-8.7\%$
against $-8.2\%$ for Wanda). Data is
what makes optimisation pay: at 32 sequences more steps hurt at every learning
rate, while at 256 every block's best held-out point lies in the last 12\% of
its budget. The curve bends (the gain per doubling falls from about 4\% to
2.7\%) and, being iso-epoch rather than iso-convergence, it bounds the floor from above
without locating it.

\textbf{What reconstruction cannot reach.} The 50\% row's remaining excess is
not in attention: block reconstruction removes 54.6\% of the one-shot excess
($r-1$) there, above the 51.1\% extrapolated from the hybrid row for an objective that
recovers nothing through attention. About a fifth is the calibration corpus
(C4 costs 0.19~nats against WikiText-2 at the one-shot tier, up from 0.12 at
32\%), and 75--80\% is the capacity of a 2:4 support in the MLPs. That is what a
dense low-rank term relaxes, so the correction, not a larger objective, is the
next stage.

\textbf{Closed-form initialisation, and why diagonal.}
SLiM's diagonal saliency weighting and the exact Gram weighting both fit $LR$ in seconds. Matched at three ranks, the
Gram form wins on held-out C4 layer error at every rank but at best ties on
WikiText-2 loss: only 28--53\% of the layer error it removes reaches the loss,
against 52--67\% for the diagonal form, because it fits the calibration
corpus's second-order structure. It also derails one or two windows in four of
its five fits (windows that open on WikiText-2 number-formatting tokens, via
the corrections of blocks 0 and 1 at the attention sink), while the diagonal
form derails none of six. The diagonal form is therefore a regulariser, not an
approximation, and it is our initialisation; a sink stress probe (hostile
openings spliced onto held-out text) flags such fits before any test window is
read.

\textbf{Training the correction.} From that initialisation, training $L$, $R$
and the surviving values jointly on Eq.~\eqref{eq:block-objective} beats the
closed form at the same rank and bytes by 4.67\% (paired SE 0.091\%, 52~SE),
on all 146 windows, flat across difficulty terciles and across positions from
4 on; only position 0, where the sink forms upstream of the MLP objective,
keeps a 0.05-nat edge for the closed form. For trained corrections a unit of
held-out block error removed reaches the loss nearly one for one, against
about one half for the closed form, which made the result predictable: the
rank-77 run's exchange rate projected 1.518 at rank 154, and 1.5166 was
measured. Rank still carries capacity (joint rank 77 does not reach closed-form
rank 154: 1.6616), and the curve is steep: joint training removes 0.141, 0.232
and 0.361 of the 0.648~nats left by reconstruction at ranks 77, 154 and 308.

\textbf{Quantizing the correction.} In 4 bits the correction's product carries
13.5\% relative error (rank 154), yet the ratio moves by only 0.38\% at rank 154 and 0.65\%
at rank 308: 98.3\% and 98.2\% of the correction's benefit survive, and INT8 is
lossless at both ranks ($+0.007\%$). The correction targets a residual it does
not fully capture, and its quantization noise enters nearly orthogonally to
that residual, adding in quadrature to a much larger term; its loss cost is
second order. Two details carry the result: the activation-weighted clip
(\S\ref{sec:recipe}) and an NF4 codebook for the near-Gaussian $L$ (9.2\% vs.\
12.0\% error for uniform INT4). The pre-registered bar, $+0.5\%$, was set at
rank 154; at rank 308 INT4 exceeds it by 0.15 points and ships as a recorded
decision, because it decodes 1.35~ms/token faster than INT8 on the M4; stated as
benefit kept, a bar that does not depend on rank, it is unchanged at 98\%.

\textbf{Against SLiM.} On Qwen2.5-1.5B the shipped row recovers 73.3\% of its
one-shot tier's log-excess; SLiM reports 51.1\% (one-shot adapters) and 69.6\%
(full fine-tuning) on OPT-1.3B \citep{slim}, a different model, so the shares
are not directly comparable. On OPT-1.3B itself (45.96\% sparsity, C4
calibration, SLiM's protocol, half our calibration budget; dense 14.625),
block reconstruction reaches 1.718 and joint training 1.3496 at rank 154 and
1.2989 at rank 205, against SLiM-LoRA's 1.27: $+2.3\%$, or 46.5\% of the
log-excess recovered against SLiM's 51.1\%. The recipe transfers; it does not
beat SLiM on SLiM's model, and rank, not joint training, is the remaining
lever there. On the M4 the OPT-1.3B BR row decodes at 24.25~ms/token,
$4.45\times$ HuggingFace eager, the product of a $3.56\times$ runtime factor
(our attention replaces HuggingFace's eager-only OPT attention,
\texttt{transformers}~4.45.1) and a $1.27\times$ sparsity factor, 97\% of the
FFN byte ceiling; the joint rank-154 row decodes at $4.02\times$.

\section{Experimental Evaluation}
\label{sec:evaluation}
\label{sec:decomposition}
\begin{figure*}[t]
\centering
\includegraphics[width=\linewidth]{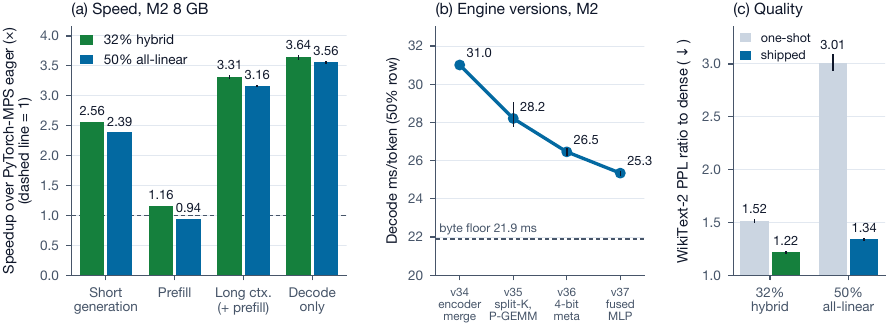}
\caption{M2, 8~GB, Qwen2.5-1.5B, v37 engine. (a) Speedup over HuggingFace
eager on PyTorch-MPS (dashed line${}=1$); bars are the mean of two runs, ticks
their range. Long context counts the 512-token prefill; decode only does not.
(b) Decode ms/token of the 50\% row by engine version (median of runs); the
dashed line is the byte floor of Table~\ref{tab:bytes} at STREAM bandwidth.
(c) Paired WikiText-2 perplexity ratio to dense (test split, 2048 context,
$\pm$SEM): one-shot tier (Wanda mask + refit) and the shipped recipe
(Table~\ref{tab:quality}).}
\label{fig:m2}
\end{figure*}

\textbf{Protocol.} Speed is decode-only ms/token (512-token prompt, 200 greedy
tokens), reproducible to 0.4\%; short generation varies by $\pm$7\% on fanless
hardware. Every run uses the production engine (native decode loop, INT8 head),
and the dense baseline runs in its own process.

\textbf{M2, 8~GB.} Both shipped rows decode $3.6\times$ faster than
HuggingFace eager (Figure~\ref{fig:m2}a): 25.3~ms/token for the 50\% row and
24.8 for the hybrid. Prefill stays near dense ($0.94\times$ and $1.16\times$),
as \S\ref{sec:bandwidth-wall} predicts. Panel~(b) shows where the 50\% row's
decode time went version by version; it now sits 3.4~ms above the byte floor,
at 87\% of STREAM. Panel~(c) shows the recipe's share: the 50\% row goes from
$3.01\times$ dense perplexity one-shot to $1.34\times$.

We price kernels in place, through allocation-free hooks
(Table~\ref{tab:kernels}), because isolated benchmarks mislead: on a single
layer, INT8 corrections beat INT4 by 0.43~ms, while in the engine INT4 wins by
0.92~ms.

\begin{table}[t]
\caption{Decode kernels priced in place through allocation-free hooks; 8-core
M2 unless noted. $\eta$ is bytes over time against the M2's 91~GB/s STREAM
bandwidth.}
\label{tab:kernels}
\centering
\small
\setlength{\tabcolsep}{4pt}
\begin{tabular}{lll}
\toprule
Kernel & GB/s & $\eta$ \\
\midrule
2:4 MLP GEMV chain      & 87.5   & 0.96 \\
Fused QKV + RoPE        & 85--89 & 0.93--0.98 \\
$o$-proj.\ + residual   & 82.5   & 0.91 \\
INT8 LM head            & 71.6   & 0.79 \\
INT4 corrections (M4)   & \multicolumn{2}{l}{0.66 ms + MB/79.6 GB/s} \\
Split-$K$ attention     & \multicolumn{2}{l}{1.2 ms per 1k tokens} \\
P-GEMM core (prefill)   & \multicolumn{2}{l}{1.86--1.89 TF, 94--96\% of MPS} \\
\bottomrule
\end{tabular}
\end{table}

\textbf{Zero-skipping does not rescue prefill.} Without sparse tensor cores, a
$2{:}4$ kernel could still multiply only the two stored values of each block
and halve its FMAs. Table~\ref{tab:zero-skip} tests this on Qwen2.5-1.5B MLP
shapes (M2, fp16 peak 2.86~TFLOP/s). Skipping half the FMAs makes the SIMT
kernel only $1.14$--$1.15\times$ faster than its dense twin at $M{=}512$: it is
limited by operand delivery (decoding positions, fetching one activation per
surviving weight), not by arithmetic. Even an ideal $2\times$ (1.52 and
1.42~TFLOP/s) stays below P-GEMM and MPS, whose \texttt{simdgroup\_matrix}
tiles reuse each operand across an $8{\times}8$ fragment (64\% and 77\% of
peak, against 27\%). A competitive zero-skipping kernel would have to close a
$2.1$--$2.2\times$ gap to our dense-tile kernel.

\begin{table}[htb]
\caption{Dense-equivalent TFLOP/s ($2MNK/t$) on an M2; the zero-skipping (ZS)
kernels do half of these FLOPs. MPS: PyTorch's GEMM on the pruned weight
stored dense. P-GEMM: our prefill kernel. ZS: a SIMT kernel doing two FMAs per
block, the activation gathered (g) or selected (s). SIMT: the same kernel on
the dense weight. All arms match an fp32 reference; median of five interleaved
rounds.}
\label{tab:zero-skip}
\centering
\small
\setlength{\tabcolsep}{3.5pt}
\begin{tabular}{lrrrrr}
\toprule
$M$ & MPS & P-GEMM & ZS (g) & ZS (s) & SIMT \\
\midrule
\multicolumn{6}{l}{\emph{gate/up}, $1536{\to}8960$} \\
64  & 1.95 & 1.47 & 0.75 & 0.52 & 0.66 \\
128 & 2.10 & 1.65 & 0.82 & 0.56 & 0.72 \\
512 & 2.20 & 1.82 & 0.86 & 0.58 & 0.76 \\
\midrule
\multicolumn{6}{l}{\emph{down}, $8960{\to}1536$} \\
64  & 1.91 & 1.26 & 0.73 & 0.49 & 0.58 \\
128 & 1.98 & 1.65 & 0.78 & 0.54 & 0.68 \\
512 & 2.15 & 1.81 & 0.82 & 0.55 & 0.71 \\
\bottomrule
\end{tabular}
\end{table}

\begin{figure*}[t]
\centering
\includegraphics[width=\linewidth]{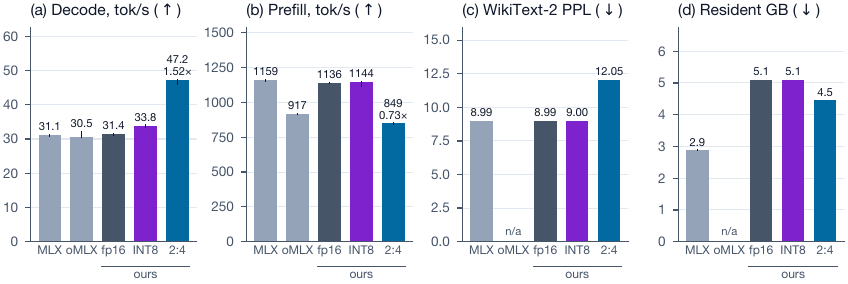}
\caption{M4, 16~GB, Qwen2.5-1.5B, one suite run of all arms (build 783dab0,
v36 kernels, i.e.\ before the fused MLP of Figure~\ref{fig:m2}b); median of
three replicates, ticks their range. Ours fp16 and INT8: our engine on the
dense weights with an fp16 or INT8 LM head. Ours 2:4: the shipped 50\% row
($2{:}4$ + INT4 corrections, INT8 head); ratios are to MLX. (a) 200 tokens
after a 512-token prompt. (c) One 20-window harness for every arm. oMLX
exposes neither perplexity nor memory.}
\label{fig:m4}
\end{figure*}

\textbf{Against MLX, M4 16~GB.} Figure~\ref{fig:m4} runs the 50\% row beside
MLX~0.32 and the oMLX~0.6.4 server. It decodes at 47.2~tok/s, $1.52\times$
MLX and $1.55\times$ oMLX. Our engine's dense arms split the gain: the runtime
alone matches MLX ($1.01\times$), the INT8 head adds $1.08\times$, and $2{:}4$
with INT4 corrections adds $1.39\times$. Dense perplexity is 8.986 in both MLX
and our engine, so the $1.34\times$ ratio has one reference. The costs are
prefill ($0.73\times$ MLX) and resident memory (4.45 vs.\ 2.88~GB, from
prefill buffers and dequantized corrections).

\textbf{Limitations.} Prefill cannot gain from 2:4 on this hardware
(\S\ref{sec:bandwidth-wall}); zero-shot task accuracy and a 4-bit MLX
baseline at matched bytes are left to the camera-ready.

\section{Hardware Discoveries: The MPS 1024-Token Defect}
\label{sec:hardware_defect}
On PyTorch 2.4.1 and macOS 13.3, an fp16 \texttt{F.linear} on MPS with 1{,}025
or more rows against the 151{,}936-row LM head returns wrong logits silently:
WikiText-2 at 2{,}048 tokens scored nonsense while 1{,}024 was correct.
Bisection isolated the single op; attention is unaffected. Every evaluation
therefore chunks the head to at most 1{,}024 rows and preflights each model
against an unfused fp32 CPU forward; no published number was affected. On the
M2 we also found that \texttt{simdgroup\_barrier} does not order a
\texttt{simdgroup\_store} against other lanes' reads (use a threadgroup
barrier), and that MPS \texttt{addmm\_} ignores the storage offset of its
output view. The rule we draw: validate every accelerated path against a
reference backend at the shapes it will run.

\section{Conclusion}
\label{sec:conclusion}
A GPU without sparse tensor cores can still profit from $2{:}4$ sparsity, as
long as the format pays through bytes rather than FLOPs and the runtime
around it is not the bottleneck. metalSPARSE streams $2{:}4$ weights at 96\%
of the M2's measured bandwidth, turns sparsity into decode time at 98\% of
its byte ceiling, and, with a runtime at parity with MLX on dense weights,
decodes Qwen2.5-1.5B $1.52\times$ faster than MLX on an M4 with $2{:}4$ on every
linear layer. A laptop-scale recipe brings that model from $3.01\times$ to
$1.34\times$ dense perplexity. Block reconstruction makes the mask choice
irrelevant, calibration data decides the one-shot result, and a correction
trained inside the block objective and stored in 4 bits recovers 73\% of the
one-shot damage. Prefill stays at parity at best, because only a sparse matrix
unit can remove FLOPs. The next steps are larger models, zero-shot task
suites, and a 4-bit dense baseline at matched bytes.


\section*{Reproducibility Statement}
To ensure the reproducibility of our experimental findings, all custom Metal Shading Language (MSL) kernel sources, C++ engine bindings, and PyTorch calibration scripts are available at \url{https://github.com/Tekraj15/metalSPARSE}. The diagonal closed-form adapter initialization is derived in Section~\ref{sec:quality_recovery}, and the full hyperparameter configuration of block-wise gradient reconstruction is given in Section~\ref{sec:recipe}. All experiments were executed on public validation splits (WikiText-2 and C4) using publicly accessible model weights (Qwen2.5-1.5B and OPT-1.3B).

\section*{Ethics Statement}
This work focuses on software-level kernel optimizations and post-training compression algorithms for Large Language Models on edge hardware. It does not involve human subjects, personal data collection, or direct harmful deployment vectors. By lowering the operational memory and power footprint required for on-device inference, this research aims to enhance user privacy by reducing dependency on centralized cloud servers.

\section*{AI Use Statement}
We disclose that generative AI models (specifically Claude Opus 4.7 and Gemini 3.1 Pro) were utilized during the preparation of this manuscript. Specifically, AI tools were used for refining prose clarity, formatting LaTeX tables, and generating template boilerplate code. All core mathematical formulations, kernel implementations, experimental data collection, hardware bug discoveries, and scientific conclusions were performed and verified independently by the authors.


\bibliography{references}
\bibliographystyle{plainnat}

\end{document}